\documentclass[12pt]{article}
\usepackage{graphicx}
\usepackage{bm}        % for math
\usepackage{amssymb}   % for math
\usepackage{color}

\def\beq{\begin{eqnarray}}
\def\eeq{\end{eqnarray}}

\def\ba{\begin{eqnarray}}
\def\ea{\end{eqnarray}}
\def\beq{\begin{eqnarray}}
\def\eeq{\end{eqnarray}}
\def\mpl{M_{\rm P}}
\def\m5{M_5}

\def\L*{{\cal L}_*}
\def\L{\mathcal{L}}
\def\({\left(}
\def\){\right)}

\def\<{\langle}
\def\>{\rangle}

\newcommand{\de}{\partial}
\newcommand{\be}{\begin{equation}}
\newcommand{\ee}{\end{equation}}

\newcommand{\bea}{\begin{eqnarray}}
\newcommand{\eea}{\end{eqnarray}}
\newcommand{\beas}{\begin{eqnarray*}}
\newcommand{\eeas}{\end{eqnarray*}}

\def\({\left(}
\def\){\right)}

\def\lsim{\mathrel{\rlap{\lower3pt\hbox{\hskip0pt$\sim$}}
     \raise1pt\hbox{$<$}}}         %less than or approx. symbol
\def\gsim{\mathrel{\rlap{\lower4pt\hbox{\hskip1pt$\sim$}}
     \raise1pt\hbox{$>$}}}         %greater than or approx. symbol

\def\lsim{\mathrel{\rlap{\lower3pt\hbox{\hskip0pt$\sim$}}
     \raise1pt\hbox{$<$}}}         %less than or approx. symbol
\def\gsim{\mathrel{\rlap{\lower4pt\hbox{\hskip1pt$\sim$}}
     \raise1pt\hbox{$>$}}}         %greater than or approx. symbol

\usepackage{amsmath}
\usepackage{amsfonts}
\usepackage{verbatim}
\usepackage{setspace}
\singlespace
\begin{document}

\begin{titlepage}

\begin{flushright}
{NYU-TH-05/18/17 }

%\today
\end{flushright}
\vskip 0.9cm

\centerline{\Large \bf A Scale-up of $\Lambda_3$}
\vskip 0.7cm
\centerline{\large Gregory Gabadadze} 
\vspace{0.1in}
\vskip 0.3cm

\centerline{\em Center for Cosmology and Particle Physics,
Department of Physics,}
\centerline{\em  New York University,  New York,
NY, 10003, USA}

\centerline{}
\centerline{}

\vskip 1.cm

\begin{abstract}

Pure massive gravity is strongly coupled  
at a certain low scale, known as  $\Lambda_3$. 
I show that the theory can be embedded into another one, 
with new light degrees of freedom, to increase  the  strong scale  to a significantly  
larger value.  Certain universal aspects of the proposed mechanism are discussed, notably 
that the coupling of the longitudinal mode to a stress-tensor is suppressed, 
thus making the  linear theory consistent with the fifth-force exclusion.  
An example of the embedding theory studied in detail is  5D AdS massive gravity, with a large 
cosmological constant. In this example the 4D strong scale can be increased by 19 orders of 
magnitude.  Holographic duality  then suggests that the strong scale of the 4D massive gravity can be increased by  
coupling  it to  a 4D non-local CFT,  endowed with a UV cutoff; however, the 5D classical 
gravity picture appears to be more tractable.

\end{abstract}

%\vspace{3cm}

\end{titlepage}

%\newpage

\section{An outline}

This article addresses the strong coupling problem 
in the 4D nonlinear diff-invariant theory of massive gravity 
\cite{deRham:2010ik,deRham:2010kj}. 
It does so by  a mechanism that  raises the strong scale
to a parametrically  larger  one, due to embedding of 
the theory into extra dimensions; the resulting theory does not seem to have 
obvious contradictions with the observations.

Among many classical solutions, Minkowski space is a solution of massive gravity, as it is 
of  the massless theory.  It is  instructive to discuss properties of massive gravity, 
especially the ones distinguishing it from the massless theory,  on the Minkowski background. 
These properties are determined by Poincar\'e invariance, and the fact that massive spin-2 representation of the 
Poincar\'e group has 5  degrees of freedom \cite {FP}. One of these 5, is the  longitudinal mode, 
properties of which are vital  to the viability of the theory \cite {vDVZ}, \cite {Arkady,DGP,DDGV},  
\cite {AGS} (see also, \cite{Luty,CedricRomb,Nicolis});  this mode will be the main  focus of the present work. 

Note that helicity is a good label  for a massive  state in the  small mass,  or equivalently, 
the high momentum approximation;  this is an approximation adopted  in the paper, and the 
longitudinal mode,  referred as $\pi$, is identified with the helicity-0 state.

The first important  property, dictated by  Poincar\'e invariance,  and the 
requirement of the absence of negative  energy states \cite {FP},  
is that the $\pi$ has no kinetic term in the linearized theory \cite {AGS}.
In the same field basis,  $\pi$  does not couple to the stress-tensor of matter.  
It has, however,  a kinetic mixing with the helicity-2. The latter can be diagonalized, to get a  
kinetic term  for $\pi$. The diagonalization induces  unsuppressed coupling  of  
$\pi$ to the matter stress-tensor. Thus, in the linearized theory 
the matter particles interact via the field $\pi$, in addition to their  interactions via the tensor field, 
leading to inconsistencies \cite {vDVZ}.  All the above properties are robust, 
guaranteed  by Poincar\'e invariance,  and the absence of any additional degrees of freedom 
beyond the 5. 

The very same requirement of  Poincar\'e invariance  dictates 
the nonlinear properties of $\pi$:  in the basis  in which  $\pi$  has no kinetic term, 
it also has no self-interaction terms in the massless limit.\footnote{More precisely, 
we'll  be using to the so-called decoupling limit: the mass goes to zero, simultaneously 
with Planck mass going to infinity, while  their certain geometric mean stays constant \cite {AGS}. }  It only 
has interactions with the helicity-2 and helicity-1 states \cite{deRham:2010ik}.  The diagonalization
that induces the $\pi$ kinetic term,  also  generates nonlinear self-interactions of 
the $\pi$ mode \cite{deRham:2010ik}.  Due to the fact that the kinetic term was absent before the 
diagonalization,  the generated self-interactions end up being suppressed by a scale, 
 \cite {AGS,Luty}, that becomes small  with smaller graviton mass, $m$, 
\beq
\Lambda_3= (\mpl m^2)^{1/3}.
\label{Lambda_3}
\eeq
The fact that this scale is low for a small graviton mass,  is a blessing and  a 
curse, at the same time: it's a virtue  for the classical theory as it enables  for an efficient Vainshtein 
mechanism \cite {Arkady,DDGV}, that suppresses  $\pi$  near physical sources, 
thus rendering  the nonlinear theory consistent with the observations  
\cite {Arkady};  such a low nonlinear scale is,  however, an obvious impediment  
for an effective low energy quantum theory.  Below $\Lambda_3$ the theory is 
weakly coupled, while  at the strong  scale an infinite number 
of new symmetry-preserving non-linear  terms could  be induced,  in the 
absence of any  specific completion by new light degrees 
of freedom appearing below  $\Lambda_3$. 

Ideally,  one would like to identify new degrees of freedom -- {\it a~ la} Higgs --  
to soften  massive gravity at the scale  $\Lambda_3$ (see, \cite{Andrew2} and 
references therein).  Ideally, these new degrees of freedom would be part of a 4D local Poincar\'e invariant 
field theory.  However, it  does not appear to be  easy  to identify such a theory
in a conventional setup. One could try to couple massive gravity to  a local  field theory  
in an unconventional way, via nonlinear terms involving St\"uckelberg fields. However, 
there are an infinite number of such couplings, and  it is not straightforward to identify 
a good principle that would help to select a finite  number of such  terms, \cite {Giorgi}.

Under these  circumstances, perhaps it then makes sense to adopt  
a  provisional approach and seek just to raise parametrically 
the strong coupling  scale, instead of striving to soften the interactions,  
{\it a~la} Higgs. If successful, such an  approach may also hint to a possible 
completion beyond the strong scale.

One could attempt to do this by introducing  a full-fledged 
kinetic term  for the $\pi$  mode; such a term would 
rescale  $\Lambda_3$ to a higher value.  For instance, 
the Vainshtein mechanism operates  due to  the kinetic terms for $\pi$ 
generated on various backgrounds \cite {Arkady,DDGV}. 
However, this cannot be done by any 4D local field theory on pure Minkowski 
background, where the lack of the $\pi$ kinetic term is mandated by Poincar\'e 
invariance, and absence of ghosts. 

Thus, one  is prompted to  think of an embedding  of massive gravity  into a theory that at  
low energies would not reduce to a local 4D theory; it would  need though to preserve  
4D Poincar\'e symmetry.  In such a theory,  the notion of a single 4D massive graviton  could only
be an approximation; fundamentally, the state that resembles  a massive graviton has to have 
distinctive features.  These very features, should also enable $\pi$ to acquire a full-fledged kinetic term, 
and raise the strong scale.  Needless to say, this theory should  be consistent with  the observations. 

I'll show that such an embedding is possible.  The larger theory is higher dimensional,   
but admits a 4D Minkowski background. The  effective 4D theory 
is that of a massive graviton coupled to an infinite number 
of gapless 4D modes.  This coupling induces a large non-local 
4D "kinetic term" for $\pi$, even though the higher dimensional theory  is local.
In the leading approximation,  the non-local 4D "kinetic term" reduces to just an ordinary large 
kinetic term for $\pi$; this leads to changing of $\Lambda_3$ to a  
new scale that can be made significantly larger. In the 5D example   considered in 
Sections 3 and 4, the  scale is increased by 19 orders of magnitude.

The lack of the  $\pi$ kinetic term was a consequence of  Poincar\'e invariance. 
Yet, I claim  the existence of a term for $\pi$  that approximates  its kinetic term.
How is this possible?  It is, since the theory is truly nonlocal 
from the 4D perspective. The  5D bulk theory is holographic 
dual to a non-local CFT that has no 4D stress-tensor  \cite{SophiaGG}. Hence, 
the mechanism of scaling up $\Lambda_3$ can be attributed to interactions  
of 4D  massive gravity with  such a non-local stuff, which is better described by 
5D classical local massive gravity.

Section 2 presents  the above arguments in detail, and puts them in a 
general  context. A reader who'd prefer  to see a concrete nonlinear 
theory, could  skip to Sections  3 and 4.

\section{Scaling up}

Let us begin with a telegraphic summary of known facts about the  origin  of (\ref {Lambda_3}), 
(for detailed discussions see \cite {AGS}, and \cite{deRham:2010ik}).
Consider  linearized massive gravity on the Minkowski background  in the limit when the 
mass tends to zero, while $\mpl$ is very large, and  only the  leading relevant 
terms are retained.  Keep track of the helicity-2 mode, denoted by $h$ (all the indexes omitted),  
and helicity-0 mode, denoted by $\pi$;  schematically,  the quadratic Lagrangian for $h$ and $\pi$   
looks as    
\beq
{\cal L}_2 = (\partial h)^2 + m^2 h \partial \partial \pi + hT\,,
\label{FP}
\eeq
where $T$ is the matter stress-tensor,  and $\mpl=1$, here and below, unless it's explicitly shown. 
The key is that $\pi$ has no kinetic term; it only 
has a kinetic mixing with $h$ \cite {AGS}.
The mixing term  is proportional to $m^2$, however, $m^2$ can be absorbed into a definition 
of $\pi$, and should not appear in physical  observables in the approximation considered.
Indeed, one can diagonalize  (\ref {FP}) by a field redefinition, $h= {\hat h} + m^2 \pi$, to get
\beq
{\cal L}_2 = (\partial {\hat h})^2  - m^4 ( \partial \pi)^2  + {\hat h} T +m^2 \pi T\,,
\label{FP1}
\eeq
and rescale,  $\pi \to  { \pi} /m^2$.
This makes the quadratic Lagrangian  (\ref {FP1}) 
independent of $m^2$, and renders the $\pi$ coupling to the trace of the 
stress-tensor as strong (or as weak) as that of ${\hat h}$. Furthermore, 
the field redefinition and rescaling  affect the nonlinear interaction terms of $\pi$ 
-- they  end up  being proportional to  inverse powers of $m$:
\beq
m^2 \pi (\partial \partial \pi)^2  \to  { {\pi} (\partial \partial { \pi})^2 \over 
 \mpl m^2}\,.
\label{cubic}
\eeq
In the last term $\mpl$   has been restored to show that the strong 
scale coincides with (\ref {Lambda_3}). The smaller the value of $m$, the lower the strong scale.  
Hence, the origin of the low strong scale is the lack of the $\pi$ kinetic term in (\ref {FP}), that would not be 
proportional to $m^4$.

Could one generate a  conventional kinetic  term  for $\pi$?   Such a term is known to be present 
on curved backgrounds \cite {Ian,Massimo}. 
For instance, on $AdS_4$,  with the cosmological constant, $-\Lambda<0$, one would obtain, 
$-{\Lambda  m^2} (\partial \pi)^2$, in addition to the  flat-space 
kinetic term,  $-{m^4} (\partial \pi)^2$, generated  after the diagonalization of the $h-\pi$ kinetic mixing   
(both kinetic terms are written here in terms of $\pi$ that has not yet been rescaled 
to a canonically normalized   field). 
This would raise the strong scale to a higher value, as long as the magnitude of the 
cosmological constant is large, $\Lambda>>m^2$.  

The very same phenomenon of  the enhancement of the kinetic term of $\pi$ on various backgrounds is 
responsible for the Vainshtein mechanism \cite{Arkady,DDGV}; the mechanism  is usually discussed in 
the context of spatially-localized sources, but it actually is universal,  and  applies to any 
background that has a characteristic physical scale.   

Ideally, one would like the strong scale to be raised in an entire space-time,  as in the example 
of $AdS_4$.  However, gravity in the observed world   cannot be approximated by $AdS_4$, or any 
other  curved space-time. Hence, such a mechanism is not immediately useful in a  theory that aims  to 
describe the  world around us. 

Nevertheless,  the above considerations suggest a path forward: if instead 
we assume that 4D massive  gravity is embedded
into a $D$-dimensional  ($D=4+n>4$) massive  gravity with a large characteristic $D$-dimensional 
curvature scale,  ${\bar \Lambda}$, then one would get a large $D$-dimensional  
kinetic term for the $D$-dimensional longitudinal  
mode, $\Pi(x^\mu, z^1,z^2,...,z^n)$,
\beq
-M_D^{2+n}{\bar m}^2 {\bar \Lambda}  \left ( \partial_D \Pi(x^\mu, z^1,z^2,...,z^n) \right )^2\,,
\label{bulk_mass}
\eeq
where $M_D$ is the higher-dimensional Planck mass, ${\bar m}$ is the higher dimensional 
graviton mass, and  $z^1,z^2,...,z^n,$  denote the extra space coordinates.  For clarity, we consider  
the case, $m\sim {\bar m}<<\sqrt{\bar \Lambda}$.  If extra space is
warped or compactified,  one gets an effective 4D description below a certain energy
scale. Thus, the  large bulk kinetic term should imply, at least in some constructions,  
a large 4D kinetic term  for the $\pi$ 
\beq
-M_D^{2+n}L^n{\bar m}^2 {\bar \Lambda}(\partial \pi (x^\mu))^2\,,~~~\pi (x^\mu)  =  \Pi (x^\mu, z^1=0,z^2=0,...,z^n=0)\,,
\label{large_pi} 
\eeq
where $L\sim {\bar \Lambda}^{-1/2}$ is the radius-curvature of the extra space, and the above expression
is valid  at distance scales larger than  $L$.

This is not enough though, one would  still  need to  obtain a (nearly) flat  4D world, 
in spite of the $D$-dimensional  space-time being curved. By no means this is automatic 
or  trivial.  Presumably,  fine-tuning  of  free parameters of the theory will be needed
to achieve this. In the 5D example considered  in the next section 
the fine-tuning is explicit.  Once this tuning is done, the goal is achieved: 
the $\pi$ would have a large 4D kinetic term determined 
by the curvature  of the embedding space, even though the  4D space is flat.
In such a theory, the strong scale  would be set 
by  the parameters of the $D$-dimensional  theory, and can 
be made much larger than $\Lambda_3$.
 
However, the above arguments appear puzzling: 
the lack of a conventional 4D kinetic term for $\pi$ was a consequence of  Poincar\'e invariance of a
local 4D field theory, coupled to gravity in a conventional way. How could this be reconciled
with the above proposal suggesting that there is  a new kinetic term for 
$\pi$ on a 4D Minkowski background?  One option to resolve the puzzle 
is for the theory to be truly nonlocal from the 4D perspective, e.g., contain  an infinite 
number of  light 4D states which cannot be repackaged into a local 4D field theory. 
Such an arrangement would evade the apparent contradiction. Hence, the kinetic term 
(\ref {large_pi})  should only be an approximation,  to a certain nonlocal 4D term reflecting 
an infinite number of the light states. 
   
Where could such states come from? Due to  the $D$-dimensional  embedding there 
will be new  degrees of freedom, that will appear as Kaluza-Klein (KK) modes from the point of view 
of the 4D effective theory. In general,  KK modes might be discrete of continuum, with or without a 
gap.  In the  present case, however, an infinite number of the KK 
states should be light,  with masses 
\beq
m_{KK} <  \Lambda_3\,,
\label{mewmasses}
\eeq
making the low energy theory to differ  from a theory of a single 
massive graviton coupled to a local 4D theory.  Furthermore, there seems to  be a 
more stringent requirement in this framework:  from the 4D perspective,  
the bulk theory cannot have a mass gap  greater than $m\sim {\bar m}$; if such  a gap existed 
the 4D  graviton could not be  a state  of  mass $m$. In the explicit  example 
considered  in Section 3, this requirement  is well-satisfied since   
there are an infinite number of light KK modes below the graviton mass scale, 
\beq
m_{KK} \lsim m\sim {\bar m}<< \Lambda_3\,.
\label{LightKKs}
\eeq
These light modes, in general,  might change the large 
distance behavior of massive gravity,  even in the  regime  of validity of the effective field theory, 
and as noted earlier, should not be representable in terms of a local 4D theory.  
These aspects should  be studied in concrete models, e.g., in the theory of Section 3.

Last but not least. The $\Lambda_3$ scale is a 
cornerstone  of making massive gravity  compatible with observations; the 
nonlinear terms,  like the one in (\ref {bulk_mass}),  lead to the Vainshtein  
mechanism through which the $\pi$ mode is suppressed near any realistic astrophysical 
source.  For this suppression to take place in an observable vicinity of any meaningful 
source, the  strong scale  should be low enough. Raising the strong scale would  confine 
the Vainshtein mechanism to shorter distances, and would  lead to contradictions with the observations. 

Luckily, in the proposed theory,  there is no problem  to start with, since the 
large kinetic term for the longitudinal mode  also leads to a suppression of  its linear coupling 
to a stress-tensor. Hence, there is no need to invoke the nonlinear  screening mechanism. 
This is straightforward to see from the  following schematic quadratic Lagrangian, motivated
by the higher-dimensional considerations of this section
\beq
{\tilde {\cal L}}_2 =  \mpl^2 (\partial {\hat h})^2 - \mpl^2 m^4( \partial  \pi)^2  -  M_D^{2+n}L^n{\bar m}^2 {\bar \Lambda} 
( \partial  \pi)^2  + {\hat h} T\,+ m^2 \pi T,
\label{FP4}
\eeq
where $\bar m$ and $\bar \Lambda$ are the bulk graviton mass, and  bulk curvature scales respectively, while 
the new kinetic term is attributed to the existence of the extra dimensions (as before, it should 
just be approximating an essentially  nonlocal term).  
As long as,  $M_D^{2+n}L^n {\bar m}^2 {\bar \Lambda}  >> \mpl^2 m^4$, 
the dimensionful coupling of the $\pi$ mode to the stress-tensor,  is proportional to 
\beq
  {(m^2 L/{\bar m}) \over \sqrt{M_D^{2+n} L^n} } \,,
\label{suppression}
\eeq
and  can  be made much smaller than the gravitational coupling, $G_N\sim 1/\mpl$.
We will illustrate the above general framework by a concrete 5D example in the next section.

\section{Warped massive gravity}

The action of a theory that realizes
the mechanism outlined above is  given in this Section.  It   
contains both 4D  and  5D parts; I'll start with the former, before specifying the latter.

The action for  the 4D metric, $g_{\mu\nu}(x),~\mu,\nu=0,1,2,3$, contains   the 4D Einstein-Hilbert term
with the coefficient $M_4^2$,  the cosmological constant, $\Lambda>0$, and the 4D dRGT mass 
term \cite{deRham:2010kj}, with the mass parameter $m$:
\begin{equation}
S_4={M^2_4} \int d^4x \sqrt{-g}\left( R(g)- 2 \Lambda +2m^2\mathcal{U} \left({\cal K} \right) \right),
\label{mGR4D}
\end{equation}
where  the  diff-invariant potential  $\mathcal{U}$  is a function of the inverse metric $g^{-1}$,  and the 
fiducial Minkowski metric, ${ \gamma}_{\mu\nu} = \partial_\mu \varphi^a 
\partial_\nu \varphi^a  \eta_{ab} ~(a,b=0,1,2,3;~ \eta_{ab} =diag (-1,1,1,1))$, in an arbitrary 
coordinate system;    this potential  can be written in  the following form \cite{deRham:2010kj,Koyama,Theo}:
\begin{equation}\label{mGRUK0}
\mathcal{U}(\mathcal{K})=\det_2(\mathcal{K})+\alpha_3\det_3(\mathcal{K})+\alpha_4\det_4(\mathcal{K}),
\end{equation}  
where  the matrix $\mathcal{K}=1-A$, and the matrix $A$  is defined  as one of the roots of,  
$A^{\mu}_{\alpha}A^{\alpha}_{\nu}=g^{\mu \alpha}\gamma_{\alpha \nu}$, so that 
$\mathcal{K}=1-\sqrt{g^{-1}\gamma}$\,,~\cite{deRham:2010kj}; $\varphi^a$ denote four scalar fields
(St\"uckelber fields).

The fiducial Minkowski metric, $\gamma$, is not dynamical.   In a certain gauge, and in the high energy limit 
the  four scalar fields, $\varphi^a(x)$,  parametrize three degrees of freedom of a massive graviton, 
helicity $\pm 1$ and  $0$.   Geometrically, these four fields  can be regarded as general coordinates  of a 
certain fiducial  4D Minkowski space-time; they guarantee the full 4D diff invariance of the theory 
(see discussions in \cite {GGKurt}).

The proposed extension of the theory is as follows:   
the bulk 5D gravity is  massive,  with the 5D mass ${\bar m}$,  and is 
endowed with a  negative 5D cosmological  constant, 
$-{\bar \Lambda}<0$. We assume that ${\bar \Lambda}>>{\bar m}^2$.  
There is a positive-tension brane in the  5D space;  its tension  is nothing but the 
4D vacuum energy density, $M_4^2\,2 \Lambda >0$, introduced in (\ref {mGR4D}). 
Moreover, the tension  will have to be tuned to $\bar \Lambda$ (in Planck 
units),  to get the flat 4D world-volume solution in the absence of any additional  brane stress-tensor, 
or brane gravity. The result is the Randall-Sundrum (RS) brane \cite {RS}.   
In spite of the mass term  in the bulk,  there is a
solution identical to the RS solution, as will  be shown below.  

The nonlinear 5D massive gravity action  for the 5D  bulk metric ${\bar g}_{MN}, ~ M,N=0,1,2,3,5$,
and 5D fiducial metric ${\bar f}_{MN}$,  takes the form 
\beq
\label{mGR5D}
S_5=\m5^{3} \int d^{4}x\,dz\, \sqrt{-{\bar g}} \left ( {\bar R}({\bar g}) + 2{\bar \Lambda}+
2{\bar m}^2 {\cal V}({\cal {\bar K}}^M_N) \right )\,,
\eeq
where 
\beq
{\cal {\bar K}}^A_{\phantom{A}B} = \delta^A_B - \sqrt{{\bar g}^{AM}{\bar f}_{MB}},~~~~{\bar f}_{MN} = 
\de_M\Phi^I\de_N\Phi^J  {\tilde f}_{IJ}(\Phi)\,,
\label{Kfid}
\eeq
and $\Phi^J(x^\mu,z),~(I,J=0,1,2,3,5),$ denote the five  scalar St\"uckelberg fields. The term ${\cal V}$ is 
the 5D dRGT potential, represented by
a  sum of all the determinants of the matrix  ${\cal {\bar K}}$, 
\begin{equation}\label{mGRU5D}
\mathcal{V}(\mathcal{{\bar K}})=\det_2(\mathcal{\bar K})+{\beta}_3\det_3(\mathcal{ \bar 
K})+{\beta}_4\det_4(\mathcal{\bar K}) + \beta_5\det_5(\mathcal{\bar K}).
\end{equation}  
The replacement  of the Minkowski fiducial metric 
by a more general one, ${\bar f}_{MN}(\Phi)$, was shown in \cite {HasanRosen} 
to retain the key property of the dRGT theory that enables it to eliminate the unwanted, 
ghostly, degree of freedom. Thus we proceed with (\ref {Kfid}).

Note that the 4D components of the 5D  and 4D metrics,  ${\bar g}$ and $g$,  the 5D and 4D St\"uckelberg 
fields, and  the fiducial  metrics,   are respectively related as follows:
\beq
{\bar g}_{\mu\nu} (x, z)|_{z=0} = {g}_{\mu\nu} (x)\,, ~~ \delta_J^a  \Phi ^J(x,z)|_{z=0}= \varphi^a(x)\,,~~ 
\delta^I_a  \delta^J_b {\tilde f}_{IJ}(\Phi)|_{\Phi^z=0} 
= \eta_{ab}\,.
\label{5D4D}
\eeq 
The full theory is specified by the above boundary conditions,\footnote{Vanishing of the classical  
value of $\Phi^z$ at the boundary, 
as implied by  (\ref {5D4D}), might  rise a concern that it could lead to vanishing of a kinetic term 
for some fluctuations at the boundary. However,  the kinetic term does not necessarily 
vanish when $\Phi^z$ does, since its  strength  is proportional to  the first  
{\it derivative} of  $\Phi^z$  at the boundary;  hence, if $\Phi^z$ vanishes  proportionally  to $z$ -- 
as it will be the case for our solutions -- no kinetic terms will vanish  at the boundary, as will be seen
in the next Section.}  
applied to the total action of the theory that reads as follows
\beq
S_{total} = S_4+S_5 + S_{GH}\,, 
\label{Stotal}
\eeq
with $S_4$ and  $S_5$  defined above;   $S_{GH}$
is the Gibbons-Hawking boundary term that  guarantees 
that the bulk equations  are those of Einstein, modified by 
the mass terms.\footnote{Note that the 5D mass  and potential terms in (\ref {mGRU5D}) do not 
contain second derivatives of $\Phi^J$, hence no new boundary terms are 
introduced for the variational procedure. However, in a non-unitary gauges 
the fluctuations on the classical background, $\delta \Phi = \Phi - \Phi^{cl}$, are often decomposed
in terms of the helicity-2,1, and 0 fields (see, the next Section).
Such decomposition introduces in the action second derivatives acting of the helicity-0 field. 
In the bulk theory such terms can  be converted into the first derivatives,  
plus total derivatives. The latter aren't important  unless there 
is a boundary, as in the present case.  Thus,  one   has to introduce
more boundary terms in the action to guarantee that the  variational principle for the helicity  
fields is well defined. Since the present  Section is dealing with the classical equations of motion 
and not with the fluctuations (i.e., no variation is taken w.r.t. the helicity fields)  I'll ignore these new 
boundary terms here (they'd  vanish on classical solutions),  but will discuss them in the next Section.}

In what follows I will regard the brane to be  the boundary of the 
5D space, at $z=0^+$, and consider the equations in that setting, as was done in 
\cite {GarrigaTanaka,Luty} (Another 
space can be glued to it according to prescribed rules; for instance, by  
postulating $Z_2$ symmetry  across the brane, as in RS, or using a more general setup, without imposing 
$Z_2$, \cite {Koyama2,GGShang}, however, this is not done here).  
Hence, one can separate the bulk and brane equations of 
motion using the general formalism of \cite {Cedric,Misao}. In the bulk, for $z>0$,
\beq
M_5^3 ({\bar G}_{AB} - {\bar \Lambda} {\bar g}_{AB}) =  M_5^3 {\bar m}^2 {\bar \Theta}_{AB}\,,
\label{bulk_eq}
\eeq
while  the   equation at the brane takes the form:
\beq
M_4^2 (G_{\mu\nu}  + \Lambda g_{\mu\nu} ) - M_5^3 (k_{\mu\nu} - g_{\mu\nu} k)
= M_4^2 m^2 {\Theta}_{\mu\nu}+ M_5^3 {\bar m}^2 [\sqrt{-{\bar g}} \,{\bar \Theta}_{\mu\nu}/\sqrt{-{g}} ]\,,
\label{junction_eq}
\eeq
where ${\bar \Theta}_{AB}$ and  ${\Theta}_{\mu\nu}$ are the stress-tensors  derived from 
the 5D and 4D  mass  terms respectively, and $k_{\mu\nu}$ denotes the value 
of the extrinsic curvature at $z=0^+$, while 
the square brackets, $[\cdots]$, denote  the boundary term obtained via 
the $z$ integration of the quantity in the brackets (which will be zero in all conventional cases).  
Variation of the action (\ref {Stotal}) with respect to the 
St\"uckelberg fields, $\Phi^J(x^\mu,z)$,  and supplied with the boundary conditions (\ref {5D4D}), 
gives the equations that are  satisfied, as long as  (\ref {bulk_eq}) and   (\ref {junction_eq}) are obeyed.   

The fiducial metric is chosen to be  the RS metric in the $\Phi$ space, with $\Phi^z \geq 0$:
\beq
ds_{Fid}^2 = {\tilde f}_{IJ}d\Phi^I d\Phi^J= {L^2 \over (\Phi^z +L)^2 } \left [\eta_{ab} d\Phi^a  d\Phi^b + (d\Phi^z)^2 \right ]\,.
\label{solutions_fiducial}
\eeq
Thus,  if we adopt  a  field configuration, $\Phi^J(x^\mu,z)\delta_J^\mu =x^\mu, ~\Phi^z(x^\mu,z)=z$, then 
the solution for the space-time metric with a flat 4D brane  located at $z=0$, does exist:
\beq
ds^2 ={\bar g}_{AB}dx^Adx^B=A^2(z) \left [\eta_{\mu\nu} dx^\mu dx^\nu + dz^2 \right ],~~~ A(z) \equiv 
{L \over z +L}\,, 
\label{solutions_metric}
\eeq
provided that the standard RS tuning between the brane and 
bulk cosmological constants is adopted, $M^2_4 \Lambda =  M_5^3 \sqrt{6 \bar \Lambda}$. 
This is so because  on the above-chosen 
field configurations, ${\bar K}^A_B=0 = {\bar \Theta}^A_B$,  
${\cal K}^\mu_\nu =0 = \Theta^\mu_\nu$, and  the equations of motion  (\ref {bulk_eq}) 
and (\ref {junction_eq}) are satisfied  due to the cancellations  between the two terms on the left hand side 
of (\ref {bulk_eq}), and between the  second, third,  and fourth terms on the l.h.s. of 
(\ref {junction_eq}). The terms that distinguish these equations from the 
RS equations  are zero on the above solution.\footnote{The above considerations suggest 
that there should be other solutions  for which neither 
${\bar {\cal K}}^A_B$ or ${\cal K}^\mu_\nu$ are zero,  and if so, then the
4D foliations might be either  $AdS$ or $dS$, as in  \cite{Kaloper,Nihei}.}

Note that instead of (\ref {solutions_fiducial}) 
one  could have started with the fiducial  metric 
\beq
ds_{Fid}^2 = {L^2 \over (\Phi^z)^2 } \left [\eta_{ab} d\Phi^a  d\Phi^b + (d\Phi^z)^2 \right ],
\label{AdS_fiducial}
\eeq
in which case one would have still obtained the solution  (\ref {solutions_metric}),
given the following relations, $\Phi^J(x^\mu,z)\delta_J^\mu =x^\mu, ~\Phi^z(x^\mu,z)=z+L$.
The latter, however, implies that $\Phi^z \geq L$, hence, the $AdS$ boundary in (\ref {AdS_fiducial}) can't be reached.
Therefore, the  two theories, one with (\ref {solutions_fiducial}), and another one with (\ref {AdS_fiducial}),
are equivalent.

Last but not least, the fiducial metric, (\ref {solutions_fiducial}),  
was introduced by "hands"  for the needs of the construction. It is however  
straightforward  to obtain as a solution of dynamical equations. For this,  one would amend  
the 5D action (\ref {mGR5D})  with the 5D Einstein-Hilbert term  for the metric, ${\tilde f}_{IJ}(\Phi)$, 
in a space-time parametrized by the coordinates $\Phi^J$, i.e., would obtain  bigravity \cite {bigravity}.  
One  would not  need to tune the Planck  scale of the second gravity, ${\tilde M}_5$,  to 
the existing one, $M_5$, however, would have to tune the  bulk  cosmological constant in  
the $\Phi$ space-time to $\bar \Lambda$
\beq
{\tilde M}_5^3 \, \int d^5\Phi \sqrt {{\tilde f}(\Phi)} \left  ( R({\tilde f}(\Phi)) + 2 {\bar \Lambda} \right )\,.
\label{bigravity}
\eeq
Furthermore, one would need to introduce a brane located 
at $\Phi^z=0$, and tune its tension to the 5D quantity, 
$2 {\tilde M}^3_5  \sqrt{6\bar \Lambda}$, as in RS.  One can then see that the solutions 
presented above -- (\ref {solutions_fiducial}) and (\ref {solutions_metric}) --  satisfy the equations  of 
motion of bigravity, for an arbitrary positive value of ${\tilde M}_5$. Thus, for simplicity, 
one can  choose the value of ${\tilde M}_5$ to be fairly large as compared to $M_5$,  to be able to 
neglect the  dynamical fluctuations of  the second metric, ${\tilde f}_{IJ}$, as it's  done  in the present work.

In the next Section we will consider  quadratic fluctuations on the background 
solution obtained  above, ignoring the fluctuations of ${\tilde f}_{IJ}$. We then 
estimate the value of the strong scale by looking at the nonlinear 
terms in the bulk,  and on the brane. The fluctuations of ${\tilde f}_{IJ}$ do not affect 
these considerations since neglecting them  neglects a massless graviton, which, unlike a massive 
graviton,  has  no strong interactions for weak sources.

\section{Estimating the new strong scale} 

The quadratic massive gravity Lagrangian in the bulk of $AdS_5$  reads as follows 
\beq
{\cal L}_{5D} =  M_5^3  \sqrt { {\bar g}^{AdS}} ~\left  (  -{\bar h}_{AB} {\bar {\cal E}}^{ACBD} {\bar h}_{CD}  -
{{\bar m}^2 \over 2} (  {\tilde h}^2_{AB} - {\tilde h}^2)\right  )  + \sqrt { {\bar g}^{AdS}} ~{\bar h}_{AB} {\bar T}^{AB}\,,
\label{FP_5D}
\eeq
where ${\bar {\cal E}}^{ACBD}$ is the Einstein operator on $AdS_5$ (see, e.g., \cite {SophiaGG}); 
note that the St\"uckelberg fields, $\Phi^J\delta_J^A= x^A + {1\over {\bar m}} V^A$, enter the Lagrangian via 
\beq
{\tilde h}_{AB} \equiv {\bar h}_{AB} - {1\over {\bar m}} \left ( \nabla _A {V}_B\, +  
\nabla _B {V}_A\,\right )\,,
\label{tildeh}
\eeq
and the covariant derivative and all the index  contractions are  defined by 
the background metric,  $({\bar g}^{AdS})$.  Furthermore, ${\bar T}^{A}_{\,~B}$ is  a 5D stress-tensor,
which will for simplicity be set to zero below.\footnote{Due to the explicit $1/{\bar m}$ factor in (\ref {tildeh}), the St\"uckelberg fields, and in particular the $\Pi$ 
mode, are normalized differently in this section as compared to the previous one, $\Pi (of~Section~2) =
\Pi (of~Section~4)/{\bar m}$.} 

The quadratic part of the 4D Lagrangian, on the other hand, reads as follows:
\beq
{\cal L}_{4D} =  M_4^2 \left  (  -{h}_{\mu\nu} {{\cal E}}^{\mu\alpha\nu\beta} {h}_{\alpha\beta}  -
{{ m}^2 \over 2} (  {h^\prime}^2_{\mu\nu} - { h^\prime}^2)\right  )  + { h}_{\mu\nu} { T}^{\mu\nu}\,,
\label{FP_4D}
\eeq
where ${{\cal E}}^{\mu\alpha\nu\beta}$ is the Einstein operator  for 4D Minkowski space, 
 $h^\prime$ is defined as 
\beq
{h^\prime}_{\mu\nu} \equiv {h}_{\mu\nu} - {1\over {\bar m}} \left ( \partial_\mu  {v}_\nu\, +  
\partial_\nu {v}_\mu\,\right )\,,
\label{hprime}
\eeq
and all the indices are contracted by  the inverse of the 4D Minkowski metric,  $\eta^{\mu\nu}$.
Furthermore, as specified in Section 3, the bulk and brane fields are related as follows:
\beq
h_{\mu\nu}(x) = {\bar h}_{\mu\nu}(x, z=0) \equiv  {\bar h}_{\mu\nu}|\,,~~~v_\mu(x) = 
V_\mu(x, z=0) \equiv V_\mu |\,,
\label{bb_connction}
\eeq
where the $"|"$ sign denotes evaluation at $z=0$. Also, note that 
$ h^{\prime}_{\mu\nu} (x) \neq  {\tilde h}_{\mu\nu}|$, due to the nonzero connection terms 
in the covariant derivatives in the expression (\ref {tildeh}).\footnote{This implies, in particular,  
that in  the unitary gauge the brane will be bent. This gauge will not be  used in the present work.} 

Before proceeding further, a comment on the boundary terms: since there are second derivatives acting 
on $\bar h$ in (\ref {FP_5D}),  the Gibbons-Hawking boundary term on the brane  
is implied,  to give the correct bulk   equations for $\bar h$. Moreover, 
the vector field $V_A$ will be  further decomposed below, introducing 
more terms in the action with second derivatives  applied to a field. 
Hence,  I'll introduce below new boundary terms  for $V_A$,  
to have the variational procedure well defined, at least in the limit specified below.

The precise limit that will  be taken is 
\beq
m\sim {\bar m} \to 0\,,~~M_5 \to \infty\,,~~M_4 \to \infty\,,
\label{limit} 
\eeq
moreover, ${\bar \Lambda}$ will be held fixed,  and it shall also become  clear below
as to why we will keep fixed the  scales,  $\Lambda_{5/2}\sim (M_5^{3/2}{\bar m})^{2/5}$, 
and $\Lambda_2\sim (\mpl {\bar m})^{1/2}$ .

A few words about the symmetries: ${\tilde h}$ is invariant under 5D linearized diffs, $\delta_d {\bar h}_{AB} = 
\nabla _A \Omega _B\, +  \nabla _B \Omega_A\,$, and $\delta_d V_C = {\bar m} \Omega_C$, where 
$\Omega_C$ is a 5D vector. Hence, the bulk action  is invariant under these transformations. The brane action, on the other hand,  
is invariant under the same transformations,  only if $\omega_z = \Omega_z|=0$. 

For now,  let us restrict the bulk diffs  by  imposing a 5D gauge in the bulk, 
\beq
{\bar h}_{\mu z} =0 ={\bar h}_{zz}\,.
\label{gauge}
\eeq    
This leaves a residual diff invariance w.r.t. the following transformations:
$\delta_d {\bar h}_{AB} = \nabla _A R_B\, +  \nabla _B R_A\,$, and $\delta_d V_C = {\bar m} R_C$,
where  the components of the 5D vector, $R_A$, are defined as follows:
\beq
R_\mu = A^2(z) \omega_\mu(x) - {L\over 2} \partial_\mu \sigma(x)\,,~~~ R_z = A(z) \sigma(x)\,,
\label{R}
\eeq
with $\omega_\mu$, and $\sigma$,  being an arbitrary 4D vector and scalar functions, respectively. 
Choosing $\sigma =0$, renders the 4D theory to be invariant under the 
residual bulk  transformations (\ref {R}). Moreover, the brane is kept fixed at $z=0$.\footnote{
To put it another way, generic  bulk field configurations for ${\bar h}_{\mu z}$  and ${\bar h}_{zz}$
can be brought to a gauge (\ref {gauge}) everywhere in the bulk  by the  diff  
transformations that would generically  entail,   $\omega_z(x) = \Omega_z| \neq0$; 
one can then use the residual $\sigma$-dependent  diff transformation (i.e., (\ref {R}), with 
$\omega_\mu=0,~\sigma\neq 0$) 
to shift,  $\omega_z \to \omega_z -\sigma$, and make, $\omega_z -\sigma=0$, by an 
appropriate choice of $\sigma(x)$.  Thus, the brane will be kept fixed at $z=0$, but 
the $\sigma$-transformations will no longer be 
allowed.}  Therefore, after  the gauge fixing,  both the 5D and 4D actions are invariant w.r.t. the following 
residual  linearized diffs:
\beq
&\delta_d {\bar h}_{\mu\nu} = A^2(z) \left (  \partial_\mu \omega_\nu  +  \partial_\nu \omega_\mu  \right ),
~~\delta_d V_\mu = {\bar m} A^2(z) \omega_\mu,  \nonumber \\
&\delta_d {h}_{\mu\nu} =  \left (  \partial_\mu \omega_\nu  +  \partial_\nu \omega_\mu  \right ),
~~\delta_d v_\mu = {\bar m}  \omega_\mu, 
\label{residual}
\eeq
with the transformations of all the other components being zero. 

The residual  diff invariance, (\ref{residual}), is the key  for counting the degrees of freedom. Generically, there are 9 
degrees of freedom carried by a 5D massive graviton. In the high momentum limit, 
the 9 can be decomposed as,  $5+3+1$, where  $5$ are carried  by a 5D helicity-2 state,  
$3$ by a 5D helicity-1,  and  $1$  by the 5D helicity-0 mode.  Given the 5D gauge choice (\ref {gauge}), 
these degrees of freedom are distributed   as follows: 5 are in ${\bar h}_{\mu\nu}(x,z)$,
while  4  remain in $V_A(x,z)$. These translate into 4D massive spin-2, spin-1, and spin-0 towers
of KK modes.  In the  limit (\ref {limit}),  the spin-2 KK tower is identical to the  RS tower \cite {RS}. 
For small nonzero masses, $m\sim {\bar m}$, the KK wave-functions would get distorted slightly, without yielding a gap,  
but turning  the RS zero mode  into a long-lived resonance, as  in a scalar theory of \cite {Rubakov};
in the case of gravity considered  here,  the zero mode needs to "eat up" three degrees of freedom 
to become a resonance. Let us see how this works:

The spin-1, and spin-0 towers, are in addition to the RS tower.  
Furthermore, the residual freedom,  (\ref {residual}), is not general  enough 
to eliminate 5D degrees of freedom, however, it could be used to remove  4D 
degrees of freedom from the 4D brane field $h_{\mu\nu}(x)$,  rendering in it only 2 (at the expense 
of keeping 3 degrees of freedom in $v_\mu$).  
On the other hand, the 4D brane field, $h_{\mu\nu}(x)={\bar h}_{\mu\nu}|$, is nothing but a linear 
superposition of all the  spin-2 KK states. To see this in the limit (\ref {limit}), 
recall  the  form of  the  KK expansion 
\beq
{\bar h}_{\mu\nu}(x,z)= \int_0^\infty  dk \,{\bar h}^k_{\mu\nu}(x) f_k(z)\,, 
\label{barhKK}
\eeq
where ${\bar h}^k_{\mu\nu}(x)$ denotes a field for a KK graviton of mass $k$, satisfying the on-shell 
condition, $\partial^\mu {\bar h}^k_{\mu\nu} = \partial_\nu {\bar h}^{k \mu}_{~\mu}$, 
while  $f_k(z)$ can be expressed via the Bessel functions. 
Then, the residual symmetry transformations, ({\ref {residual}), can be used to remove degrees of freedom 
from  a linear  superposition  state of  the spin-2 KK modes 
\beq
{h}_{\mu\nu}(x)= \int_0^\infty  dk \,{\bar h}^k_{\mu\nu}(x) f_k(0)\,,
\label{hKK}
\eeq
by imposing on it a further 4D gauge fixing condition of one's choice.\footnote{Once the KK solutions and the 
expansion (\ref {hKK}) are adopted, the 4D gauge fixing should be consistent with the on-shell condition,
$\partial^\mu {h}_{\mu\nu} = \partial_\nu { h}^{\mu}_{~\mu}$. 
Using (\ref{residual}) in the latter, one finds the allowed
residual  transformations, with $\omega_\mu$ satisfying,
$\partial^\mu \omega_\mu=0$, and  $\square_4 \omega_\mu =0$;
the  solutions of the latter  two equations enable one 
to remove 3 on-shell massless degrees of freedom from $h_{\mu\nu}(x)$. 
Away from the limit (\ref {limit}), the above KK relations get modified
by small graviton mass corrections, however, the symmetry (\ref {residual}) 
remains, and still should enable one to remove 3 degrees of freedom from $h_{\mu\nu}(x)$.}

Based on the above symmetry and gauge fixing considerations, 
the KK spectrum, away from the limit (\ref {limit}), should host one special 4D 
collective massive state,  described by  the fields, 
$h_{\mu\nu}(x)$ (2 degrees of freedom) and $v_\mu(x)$ (3 degrees of freedom), in analogy with 
the scalar field of \cite {Rubakov}.  As noted, in the limit (\ref {limit}), this special state is massless.
Its tensor part, described by $h_{\mu\nu}(x)$, 
is nothing  but the RS zero mode, carrying 2 degrees of freedom; its vector  part, $v_\mu$, 
decouples from the tensor part in the limit (\ref {limit}), 
carrying 3 degrees of freedom. Away from the limit ({\ref {limit}), 
the collective state  can be thought 
as the RS zero mode that has "eaten up" 3 degrees of freedom of $v_\mu$ to become  
massive, and thus acquired a width to decay  into the lighter KK modes. 
Hence, the tensor part of this metastable state would give leading interactions similar to 
those in  the single-brane RS model, at distances  
greater than $L$ and smaller than the inverse  graviton mass. Furthermore, the strongly coupled 
sector is due to this resonance. The helicity-2 part 
of it is weakly  interacting for weak sources,  however, the helicity-1 and helicity-0, 
can get strong. Given the bulk and brane gauges discussed above, the  helicity-1 and helicity-0 
of the resonance reside in the vector $V_A$, and its pullback,  $v_\mu$.  For these reasons, I will   focus below
on the  vector fields, and their strong coupling.

It  follows from (\ref {FP_5D}) and (\ref {tildeh}), that the 5D vector   field, $V_A$,  acquires  the Maxwell kinetic term, 
as well as a mass term,  due to the background curvature. In the limit (\ref {limit}), the vector is decoupled from
the tensor  perturbation, $\bar h$,  and its Lagrangian is proportional to
\beq
M_5^3 \sqrt{ {\bar g}^{AdS}}\, \left (  -{1\over 4} F_{AB}^2 - {1\over 2}  M_V^2   {V}_A^2 \right ) \,,
\label{KinV}
\eeq
where $M_V^2 = 4{\bar \Lambda}/3=8/L^2$.  The helicity decomposition of the 
massive vector  makes sense  for the momenta higher than $M_V\sim \sqrt{\bar \Lambda}$. 
On the other hand, one is interested in the 4D theory, that emerges in the opposite limit, when the 
momenta are 
much smaller than $\sqrt{\bar \Lambda}$. In such a regime helicity  appears to be an inappropriate  
label.  However, due to $AdS_5$ warping, a nonzero $M_V$ does  not generate  
a mass gap in the KK spectrum \cite {Rubakov}; thus,  from  a 4D perspective 
we expect a state with mass of the order of, $m\sim {\bar m} << {\bar \Lambda}$.  
If so,  then  4D helicity should be a good  label for  the momenta above  the scale of, 
$m\sim {\bar m}$, and below that of $\sqrt{\bar \Lambda}$.
Hence, the aim will be  to use  a 5D formalism that would  lead 
to the 4D decomposition  of the 5D vector in terms of the 4D helicity-0 and  helicity-1 states.\footnote{
Note that there  is a total derivative  term  en route  from  (\ref {FP_5D}) to (\ref {KinV}). 
This total derivative  induces a nonzero 4D surface term. I  introduced   a new  
boundary term in the action, proportional to, $\int\, d^4x\,(V^A\nabla_A  V_z  - V_z \nabla^C V_C)|$, to 
cancel  the surface term generated by the total derivative.  This very boundary term 
removes some of the induced  surface terms  for the  field $\Pi$, for which the bulk 
action  contains terms with second derivatives acting on $\Pi$; thus, due to the  introduced 
boundary term, the quadratic action for $\Pi$, in the limit (\ref {limit}),  
will contain only its first derivatives, see below.}  

With the above goal in mind, one can decompose the vector into its transverse and longitudinal parts, 
as follows
\beq
{V}_A = V^T_A + \nabla_A \Pi \,, ~~~ {\rm with}~~~      \nabla^AV^T_A=0\,.
\label{VT}
\eeq
The  actions for $V^T$ and $\Pi$  separate from one another, and  
the substitution of (\ref {VT}) into (\ref {KinV}) generates a kinetic 
term for  $\Pi$  proportional to the background curvature
\beq
-{M_5^3  M_V^2 \over 2} \sqrt{ {\bar g}^{AdS}}\,(\nabla_A \Pi)^2\,.
\label{KinPi}
\eeq
To reiterate, from a 5D perspective $\Pi$ is a helicity-0 mode only at very high energies, 
above $\sqrt{\bar \Lambda}$, but not  at low energies. From a 4D perspective 
the spectrum  of (\ref {KinPi})  is gapless and continuous. Most importantly, 
it hosts a localized zero mode \cite {BorutGG}. This mode is a scalar analog 
of the localized RS helicity-2 state (the helicity-1 is not localized). 
Understanding of its dynamics will be crucial.

The following rescaling makes the above kinetic term canonically normalized
\beq
\Pi  = { { \Pi}^c \over  \sqrt{ M_5^3  M_V^2}}\,.
\label{rescale}
\eeq
Furthermore, the term, (\ref {KinPi}),  would appear in the  theory away from the limit 
(\ref {limit})  in addition to the kinetic mixing  between $\bar h$ and $\Pi$, which 
is proportional to $M_5^3 {\bar m}$.  
After the rescaling, (\ref {rescale}),  the mixing term is proportional to $M_5^{3/2} ({\bar m}/\sqrt M_V)$. 
Since one already has the kinetic terms for both the helicity-2 (proportional to $M_5^3$)  
and helicity-0 (canonically normalized), then such a mixing  can be neglected  as long as 
one stays in  the  regime, ${\bar m}<< M_V =  \sqrt{4 {\bar \Lambda}/3}$ (hence, the mixing  
vanishes in the limit (\ref {limit})). 

Note that in the basis of the fields used  here there is no  direct coupling  of the helicity-0  to the 
5D external   stress-tensor.  Such coupling would arise  due to  the diagonalization of the kinetic 
mixing term; however, the latter is  negligible,  as was just shown. 

Let us now look at  the consequence of the split  (\ref {VT}) on the brane, and in particular, 
see what  it implies for    the respective 4D decomposition 
\beq
v_\mu = v_\mu^T+ \partial_\mu \pi\,,
\label{vt}
\eeq
where $v_\mu^T = V_\mu^T|$, $\pi =\Pi|$; note that $v_\mu^T$ is not a 4D transverse 
vector;  in fact, from the 4D perspective, it's unconstrained,  and is determined by the effective  
4D dynamics.  One would like to understand   the meaning  of (\ref {vt}) when substituted into the   
4D mass term:
\beq
- {M_4^2 m^2\over 2 {\bar m}^2} \left ( ({\bar m}h_{\mu\nu} - \partial_\mu v_\nu -  \partial_\mu v_\nu )^2 
- ({\bar m}h^\mu_\mu  -  2 \partial^\mu v_\mu )^2 \right )\,.
\label{4Dmass}
\eeq
To clarify  the  role  of (\ref {vt}) 
in 4D, one should point out  that the decomposition (\ref {VT}) is arbitrary up
to the following "gauge" transformations, $\delta_g V_A^T = \nabla_A S\,,  \delta_g \Pi =  - S$, where 
the scalar $S$ satisfies, $\nabla^2 S=0$. The latter equation, in the presence of the 
brane,  has a non-trivial decaying  solution 
\beq
S(x,z) = \left (  {z+L\over  L} \right )^2 {K_2((z+L)\sqrt{-\square_4}) \over K_2(L\sqrt{-\square_4}) }\,s(x)\,,
\label{A}
\eeq
where $s(x)$ is an arbitrary 4D scalar field. The above "gauge" symmetry  
can be used to move the description of the 4D degrees of  freedom between the 4D fields, 
$v^T_\mu = V^T_\mu|$ and $\pi =\Pi|$.  Indeed, under the
"gauge" transformation,  $\delta_g v^T_\mu  =  \partial_\mu s\,,  \delta_g \pi =  - s$, where $s=S|$.
Thus, one  can use this freedom to remove the longitudinal part from $v_\mu^T$, rendering  it 
only  with its transverse part, while keeping the longitudinal field in $\pi$. This appears to be a logical choice,  
since the progenitor of $\pi$, the $\Pi$ field, propagates the bulk longitudinal mode in a very high 
momentum limit, above $\bar H$.  Having this done, from now on I focus on $\Pi$, and  $\pi =\Pi |$; since the  separation of the degrees of freedom between $v^T$  and $\pi$ at low energies is not unique,  then focusing on the $\Pi$-sector, while ignoring $V^T$,  should be enough to estimate the lowest strong scale both in the bulk and on the brane.
\footnote{This is not to imply that all the vectors at low energies should be put to zero, but only  
that $\Pi$ is expected to give a lowest strong scale. In particular, in the present setup, 
$v_z$  and $\pi$ get  related due to, $\nabla^C V_C =0$, and the 
gauge choice on the brane that led to  $v^T\to v^t$.}

Hence, I turn to  the nonlinear terms for $\Pi$ in the bulk and estimate  
their strong scales.   The representative terms that manifest strong interactions in 5D,
and contain both helicity-2 and helicity-0, are proportional to 
\beq
 M_5^3  {\bar m}^2  {\bar  h} \left (  \left ({ \nabla  \nabla \Pi \over {\bar m}} \right )^2 +  
 \left ({ \nabla  \nabla \Pi \over {\bar m}} \right )^3+  \left ({ \nabla  \nabla \Pi \over {\bar m}} \right )^4  \right  )\,.
\label{bulk_dim_7}
\eeq
There  are  relative, order one  parameters  between the three terms in the parenthesis, 
however I will  omit them here and below for simplicity of presentation. 
In addition, there are also nonlinear terms containing only $\Pi$'s; these terms would have been 
total derivatives on a flat background,  however,  on $AdS_5$ they turn into full fledged terms 
in the action due to nonzero commutators of  the covariant derivatives.\footnote{This  still leaves  
total derivatives, which in the present case would induce nonzero surface terms. 
As was done in the quadratic action, I  invoked nonlinear  boundary terms  to cancel the surface 
terms; this makes the variational problem  for $\Pi$ well-defined, at least in the limit (\ref {limit}), 
where all the mixing terms vanish. In general,  
for each of the four potential  terms in (\ref {mGRU5D}),  there is a nonlinear 
total derivative term  for $\Pi$,  that contains second derivatives of $\Pi$ \cite {deRham:2010ik}; 
all of these terms will induce surface terms. The latter are  cancelled  by invoking  
the new boundary terms,  written in terms of $V$,  as discussed above.}  
Schematically, they  look   as follows:
\beq
M_5^3 {\bar m}^2 {\bar \Lambda}  \left (    \left ({\nabla \Pi \over {\bar m}} \right )\left ({ \nabla \Pi \over {\bar m}} \right )\left ({ \nabla  \nabla \Pi \over {\bar m}} \right )     + ... + \left ({\nabla \Pi \over {\bar m}} \right )\left ({ \nabla \Pi \over {\bar m}} \right )\left ({ \nabla  \nabla \Pi \over {\bar m}} \right )^3 \right )\,.
\label{CovGal}
\eeq
After the rescaling to the canonically normalized  fields,  as in
(\ref {rescale}), and using, ${\bar h} = {\bar h}^c/M_5^{3/2}$, one gets  for the nonlinear mixing  
terms 
\beq
\Lambda_{7/2}^{7/2}  \,  {\bar  h}^c  \left (  {(\nabla  \nabla \Pi^c)^2  \over  (M_5^{3/2}{\bar m}  
\sqrt {\bar \Lambda})^2  }   +  {(\nabla  \nabla \Pi^c)^3  \over  (M_5^{3/2}{\bar m}  
\sqrt {{\bar \Lambda}})^3 }  +  {(\nabla  \nabla \Pi^c)^4  \over  (M_5^{3/2}{\bar m}  
\sqrt {{\bar \Lambda}})^4 }     \right ) \,,
\label{bulk_dim_7_can}
\eeq
and for the pure $\Pi$ terms 
\beq
{(\nabla \Pi^c) (\nabla \Pi^c)  (\nabla  \nabla \Pi^c) \over  
(M_5^{3/2}{\bar m}  \sqrt {\bar \Lambda})  } +  {(\nabla \Pi^c) (\nabla \Pi^c)  (\nabla  \nabla \Pi^c)^2 \over  
(M_5^{3/2}{\bar m}  \sqrt {\bar \Lambda})^2}  + {(\nabla \Pi^c) (\nabla \Pi^c)  (\nabla  \nabla \Pi^c)^3 \over  
(M_5^{3/2}{\bar m}  \sqrt {\bar \Lambda})^3  }   \,,
\label{CovGal_resc}
\eeq
where, $\Lambda_{7/2} = ( M_5^{3/2} {\bar m}^2)^{2/7}$,  is what would have been  
the strong scale  of the  5D theory if it had no bulk cosmological constant.  However,
due to  the large bulk CC, ${\bar \Lambda} \equiv {\bar H}^2 >> {\bar m}^2$,   
the strong scale is higher; its lowest value is obtained from the terms containing  the 
$\Pi$ self-interactions only,  (\ref {CovGal_resc});  hence, 
the mixing terms (\ref {bulk_dim_7_can}),  and  the respective 
boundary terms they'd call for,  can then be ignored. The strong 
scale reads:
\beq
\Lambda_{5D} \simeq  (M_5^{3/2}{\bar m}  
{{\bar H}})^{2/7}= \Lambda_{7/2} \left ( {\bar H} \over {\bar m} \right )^{2/7}>>\Lambda_{7/2}\,.
\label{5Dhigh_scale}
\eeq

Having the 5D strong scale estimated, let  us  turn to the respective 
effective 4D theory, with the goal to estimate  
its strong scale.   The 4D description should be valid  at energies  below the scale of curvature 
of 5th dimension, $E\lsim\sqrt{\bar \Lambda}= {\bar H}\sim  L^{-1}$. 

As was already noted,  the spectrum of the KK modes   has no mass gap,  and is continuous,  
in spite of the nonzero bulk and brane mass terms \cite {Rubakov}.  Nevertheless,  for the 4D 
distance scale $r$, such that  $L<<r$, it is possible to argue that the 5th dimension  can  
be integrated out approximately. To show this, one  notes that the $\Pi$ equation in the bulk 
that follows from (\ref {KinPi}), $\nabla^2 \Pi=0$, can be solved with the decaying boundary 
conditions at $z \to \infty$
\beq
\Pi (x,z) =  \left (  {z+L\over  L} \right )^2 {K_2((z+L)\sqrt{-\square_4}) \over K_2(L\sqrt{-\square_4}) } \,  \pi(x)\,.
\label{Pi_sol}
\eeq
Then, the 4D "kinetic term" for $\pi$ can be obtained by 
substituting (\ref {Pi_sol}) into (\ref {KinPi}); as is well known,  
this leaves only a surface  term proportional to, $({\Pi} \partial_z {\Pi} )|_{z=0}$,
which in its turn gives rise to the following  term in 4D  
\beq
-{M_5^3 M_V^2  \over 2} \,\pi (x) \, \sqrt{-\square_4 }\,
{K_1(L\sqrt{-\square_4}) \over K_2(L\sqrt{-\square_4}) } \,\pi (x)\,.
\label{Kinpi_NL}
\eeq
This nonlocal term appears in the 4D effective  
Lagrangian as a "kinetic term" for $\pi$. It defines the $\pi$ propagator,  that has 
a gapless continuum of poles. These poles correspond to a gapless 
continuum of  4D particles.  

In the leading approximation for the 4D effective description,  when $L\sqrt{-\square_4}<<1$, one  expands 
the McDonald functions and finds that the $\pi$ kinetic  term is proportional  to
\beq
L \, {M_5^3 M_V^2  \over 2} \, \pi (x) \, \square_4 \,  \pi (x)\,.
\label{Kinpi}
\eeq
Thus, in the leading approximation, the scalar  in the $AdS_5$ background 
"feels" its ambient space  as if  it were of a physical size $L$, \cite {BorutGG}.

Due to the large induced kinetic term (\ref {Kinpi}), the 4D dynamics of $\pi$ should be expected 
to differ significantly from that in pure 4D massive gravity. To understand those differences 
let us look at other 4D terms containing $\pi$. One of them is a kinetic mixing term 
between the tensor, $h$,  and $\pi$
\beq
{M_4^2 m^2\over {\bar m}} h_{\mu\nu} (\partial^\mu \partial^\nu \pi - \eta^{\mu\nu} \partial^2 \pi)\,.
\label{branemixing}
\eeq
 After rescaling  (\ref {rescale}), the brane mixing term  ends up being proportional to the following ratio, 
$ q={(M_4^2 m^2/ \sqrt{ M_5^3 {\bar m}^2 {\bar \Lambda}})}\,.$
Since,  ${\bar m} \sim m << \sqrt {\bar {\Lambda}}$, and  $M^2_4 \lsim  M^3_5/\sqrt{{\bar \Lambda}}$, 
we conclude that  $q\sim {\cal O}(m)$ and, therefore, the brane mixing term can also be neglected as 
compared to the induced 4D kinetic term (\ref {Kinpi}). 

Last but not least,  there are also 
genuine 4D non-linear  terms involving  $h$ and $\pi$ \cite {deRham:2010ik}, and one would like to  estimate  
their strong scale,  in the presence  of (\ref {Kinpi}). To this end, one collects the following 
representative  linear and nonlinear terms  of tensor-scalar  and scalar-scalar interactions
\beq
- L M_5^3 M_V^2   (\partial \pi)^2 + ({M_4^2 m^2} + LM^3_5{\bar m}^2 ) {h} \left ( \Sigma^3_{n=1}
 \left ({\partial   \partial  \pi \over {\bar m}} \right)^n  \right  )+ 
{L M_5^3 M_V^2 \over {\bar m}} (\partial \pi)^2
(\partial \partial \pi)^2\cdots
 \label{dim_7}
\eeq
where the order one coefficients between the terms in the above schematic 
expression have been ignored.\footnote{The bulk cubic Galileon gives zero on the lowest order bulk 
equations of motion for $\Pi$, and hence was not included in (\ref{dim_7}).}  
Since,  $m\sim {\bar m}<<{\bar H}=\sqrt{\bar \Lambda}$,  one can obtain 
the following expression for the canonically normalized  $\pi$ field  
\beq
 -(\partial \pi^c)^2 + ({M_4^2 m^2} + M^3_5L{\bar m}^2 ) { h}  \left ( {  (\partial \partial \pi^c)   
 \over  \Lambda^{3}_{*}   } + {  (\partial \partial \pi^c)^2    \over  \Lambda^{6}_{*}   } +   
{ (\partial  \partial  \pi^c)^3  \over  \Lambda_{*}^9 } \right )  +  { (\partial \pi)^2
  (\partial \partial \pi)^2   \over  \Lambda^6_{*} }+... \,,
\label{dim_7_c}
\eeq
where, $\pi^c \equiv \sqrt{L}\,\Pi^c|$, and 
\beq
 \Lambda_{*} \simeq (M_5^{3/2} {\bar m} {\bar H}^{1/2})^{1/3}\,, 
\label{4Dhigh_scale}
\eeq
is the lowest  strong scale of the 4D theory due to the $\pi$ self-interactions. 
Note that the 4D Planck mass is determined by two contributions,   
proportional to $M_4^2$,  and $LM_5^3 $, respectively; if  for simplicity we assume the latter 
is greater than the former, then, it would follow that $M_5^{3/2} \sim \mpl {\bar H}^{1/2}$,
and, 
\beq
\Lambda_{*} \sim (\mpl {\bar m} {\bar H})^{1/3} = (\Lambda_2^2 {\bar H})^{1/3}. 
\label{scale2}
\eeq
To estimate the numerical value of this scale,  let us set, 
${\bar H} \sim 10^{16}\,GeV$,  $M_5 \sim 10^{18}\,GeV$, and ${\bar m}\sim m \sim 10^{-42} \,GeV$,
then the 5D strong  scale, $\Lambda_{5D} \sim GeV$, while the 4D strong scale  is  lower, 
$ \Lambda_{*}\sim MeV$.  The latter, however, is some 19 orders of magnitude  
greater than the strong scale of pure 4D massive gravity, 
$\Lambda_3 \sim 10^{-19}\,MeV$. \footnote{One could  also wonder if the theory can transition  to the 5D regime before
reaching the 4D strong scale (\ref {4Dhigh_scale}). That is possible if, $\Lambda_* \gsim  {\bar H}$, 
or  expressed differently,  $M_5^3 {\bar m}^2 \gsim {\bar H}^5$; the latter condition  can  
be rewritten as, $\Lambda_2 \gsim {\bar H}$, if  $LM_5^3 > M_4^2$. Since $\Lambda_2$ is of the 
order of $10^{-3}~eV$, the  effective size of such a dimension would  be 
a millimeter or larger. In that case, the strong coupling of the theory would be given by the 5D scale, (\ref {5Dhigh_scale}), which then can be estimated to be, $\Lambda_{5D} \sim 10^{-3} eV$;  
not much of a gain.} 

A few important comments are in order.  Only a simple setup was studied 
above with  just one  mass scale on the brane and in the bulk, $m\sim {\bar m}$; 
however, it is straightforward to see  that all the results above apply to the case, 
$m << {\bar m}$, and in particular to, $(m /{\bar m}) \to 0$. This is so 
since the 4D massive theory of Section 3  is perturbatively continuous in the $m\to 0$ limit 
due to the bulk physics with ${\bar m} \neq 0$,  and the scale of non-linear interactions, (\ref {scale2}),  is  
independent of $m$, in the leading approximation.

The estimate for the new strong  scale, (\ref {4Dhigh_scale}),   was  made
above assuming generic values  for the parameters, $\alpha_3,\alpha_4,
\beta_3,\beta_4,\beta_5$,   in the 4D and 5D  potentials,
(\ref {mGRUK0}) and (\ref {mGRU5D}).  However,  for certain specific  relationships between  
some of these parameters there might be cancelations  between at least some of the 
nonlinear terms, in analogy with the cancellations in  a 4D  flat space  case 
\cite{deRham:2010ik}; if so, it is then conceivable that in those special 
cases the strong scale  might perhaps be  higher than (\ref {4Dhigh_scale}). 

Furthermore, none of  the calculations above would have changed if  one ignored the 
4D Einstein-Hilbert (EH) term, but kept a fixed brane tension, $2 M_4^2 \Lambda $.   
However, the 4D EH term  might be useful for more general solutions, and would in any case  be induced 
by quantum loops in an effective theory,  even if it was not  introduced in the classical 
theory to begin with \cite {DGP}. Hence, it was included for generality.    

As a final comment, the 5D massive gravity on $AdS_5$ with the 
$AdS_5$ fiducial metric,  was argued \cite {SophiaGG}  to be holographic  dual to a theory of 
unparticles \cite {Unparticles} -- a certain non-local CFT that has no conserved 4D stress-tensor. 
If so, then the mechanism of the present work could be thought in terms of  the  CFT with a UV cutoff: the $\pi$ 
mode acquires a large kinetic term due to  coupling of  massive gravity to the 4D non-local CFT,
while the existence of the RS brane  translates into the existence of
a UV cutoff  of the 4D  CFT. Since  such a CFT appears to be  pretty exotic, the 5D
classical gravity description, used in this work, seems to be a simpler option.  However, certain aspects 
might be clearer in the CFT;  for instance, in the $m=0, {\bar m}\neq 0$ theory, 
4D massive graviton is a state that should perhaps better be viewed as a resonant spin-2 
state emerging entirely from the CFT.

\section*{5  An outlook}

Calculations  in Section 3 suggest that the theory proposed in this paper does admit   a
self-accelerated solution,  for some  values of the  parameters (for a review of self-acceleration 
in massive gravity and its extensions, see \cite {Mukohyama}, and references therein).
The main features  of the helicity-0 mode  -- that its  coupling  to an external stress-tensor 
can be  ignored, and that its strong scale is high -- would remain valid on the self-accelerated background. 
However, the quadratic fluctuations on the background should be expected to receive additional terms 
as compared to pure massive gravity. It would be interesting to see if these solutions exhibit healthy fluctuations.

Furthermore, one can straightforwardly extend the theory in various directions, for instance, 
by introducing  a  scalar field that sets the mass scale, thus providing a dynamical mass generation; 
or introduce a more restricted scalar based on dilatation  symmetry,  as in the  quasidilaton theory \cite {Quasidilaton}, and its generalizations \cite {NQuasidilaton}. One could  study self-acceleration in the warped  
version of bigravity of \cite {bigravity},  discussed in Section 3, when the metric ${\tilde f}(\Phi)$  becomes  
dynamical  due to   a 5D Einstein-Hilbert term  integrated over the 5D invariant 
volume in the $\Phi$ space-time,  together with the cosmological constant, 
$\int d^5\Phi \sqrt {{\tilde f}(\Phi)} ( R(\tilde {f}(\Phi)) + 2 {\bar \Lambda})$, and a "brane" in the $\Phi$-space is also included.

It remains to be seen if the mechanism proposed in this work  may or may not be understood 
as softening of the strong $\pi$ amplitudes by the light modes of the non-local CFT, which by itself 
should  have a strong coupling. Last but not least, would be interesting to study extensions of the theory beyond 5 dimensions, including unconventional ones along the lines of \cite {MasterTheory}, with the goal to perhaps raise the strong scale even further, ideally toward the Planck scale.

\subsection*{Acknowledgements}

I'd like to thank Borut Bajc, David Pirtskhalava,  and Takahiro Tanaka, for useful communications,
and Evelina Steponaityte,  for editorial  help. The work was supported in 
part by NSF grant PHY-1620039.

%\newpage

\end{document}